# L'intelligence artificielle générative et la marginalisation des savoirs minoritaires dans l'enseignement supérieur. Le cas du handicap

## Fatiha TALI OTMANI

**Université Toulouse Jean Jaurès-UMR EFTS**

**Résumé**

L'intelligence artificielle générative transforme l'enseignement supérieur et reconfigure les modalités de production comme de validation des savoirs scientifiques. Ces dispositifs ne sont pas neutres et participent activement à la marginalisation d'épistémologies non hégémoniques. Nous mobilisons les apports des sciences de l'éducation et de la formation, des études critiques des technologies et des *disability studies*. Les corpus d'entraînement, majoritairement anglophones et occidentalo-centrés, reconduisent une colonialité épistémique. La situation des personnes en situation de handicap en témoigne particulièrement. Les architectures technologiques les assignent à des stéréotypes ou les excluent de leur conception, ce qui produit une double marginalisation. Nous étudions comment une hybridation chercheur/IA pourrait préserver la pluralité épistémique, tout en prenant en compte les limites structurelles propres aux tentatives de stratégies palliatives de correction algorithmique.

**Abstract :**

Generative artificial intelligence redefines higher education by restructuring the processes through which scientific knowledge is produced and validated. These systems are not neutral; they actively contribute to the marginalization of non-hegemonic epistemologies. This research draws upon educational sciences, critical technology studies, and disability studies to demonstrate that training datasets, which remain predominantly Anglophone and Western-centric, reinforce epistemic coloniality. The situation of persons with disabilities provides a particularly clear illustration of this phenomenon. Technological architectures frequently confine these individuals to reductive stereotypes or exclude them from the design process, leading to a double marginalization. This article examines whether a hybridization between the researcher and the machine might preserve epistemic plurality, while acknowledging the structural limitations inherent in algorithmic correction when used as a purely palliative strategy.

## Introduction

Depuis la diffusion massive des grands modèles de langage (Large Language Models, LLMs) en 2022, l'intelligence artificielle générative (IAG) modifie les pratiques de production des connaissances, de rédaction scientifique, d'évaluation par les pairs et d'accompagnement pédagogique mobilisées par les enseignants-chercheurs. Méta-analyses et revues systématiques se multiplient pour en rendre compte (Khalifa & Albadawy, 2024 ; Mustafa et al., 2024 ; Zawacki-Richter et al., 2019).

Au-delà des discours d'optimisation s'observe un mouvement moins visible. Les systèmes d'IAG ne sont pas neutres. Ils reproduisent et amplifient des inégalités préexistantes dans la production du savoir. La composition des corpus, les choix algorithmiques de leurs concepteurs et la logique probabiliste des modèles favorisent certaines formes de connaissance au détriment d'autres. L'homogénéisation du discours scientifique qui en résulte fragilise la diversité épistémique, condition de la vitalité intellectuelle d'une communauté savante.

Cet article aborde l'enjeu à partir de deux interrogations. Par quels mécanismes l'usage de l'IAG dans l'enseignement supérieur contribue-t-il à marginaliser les savoirs minoritaires ? Sous quelles conditions une hybridation entre pratiques de recherche et outils d'IA pourrait-elle préserver, voire renforcer, la pluralité épistémique ? La littérature mobilisée couvre les sciences de l'éducation et de la formation, les études critiques des technologies, l'éthique algorithmique et la philosophie des sciences. Le propos progresse en quatre temps. La standardisation du savoir académique se trouve articulée à la colonialité épistémique. Les mécanismes d'amplification des inégalités sont ensuite examinés au prisme des biais documentés dans les systèmes d'évaluation par LLMs. Le cas des personnes en situation de handicap éclaire une invisibilisation spécifique. Les conditions d'une hybridation chercheur/IA respectueuse de la diversité des savoirs ferment la réflexion.

## L'IAG et la transformation du savoir académique. Standardisation et colonialité

### *La standardisation algorithmique du discours scientifique*

L'effet le mieux documenté de l'intégration des outils d'IAG dans la recherche académique tient à la modification du registre lexical et stylistique des publications. Cette évolution témoigne d'une transformation des conditions mêmes de production du savoir.

Kobak et ses collègues (2025) ont analysé quinze millions de résumés biomédicaux indexés par PubMed entre 2010 et 2024. Au moins 13,5 % des résumés publiés en 2024 ont été produits avec l'aide d'outils d'IAG, ce taux culminant à 40 % dans certains sous-corpus disciplinaires. L'impact sur le vocabulaire biomédical surpasse celui d'événements aussi marquants que la pandémie de Covid-19. Liang et ses collègues (2025) obtiennent des résultats convergents sur un million d'articles publiés dans ArXiv et Nature. La pénétration s'avère particulièrement forte dans les sciences informatiques (22 %), nettement supérieure aux mathématiques et aux sciences naturelles (9 %).

La standardisation linguistique observée révèle l'imposition de schèmes rédactionnels issus de corpus déséquilibrés, massivement anglophones et ancrés dans des orientations épistémologiques occidentales (Messeri & Crockett, 2024 ; Roe, 2024). Les travaux produits dans d'autres langues, à partir de traditions de pensée distinctes ou selon des méthodologies moins compatibles avec le traitement algorithmique, se trouvent mécaniquement sous-représentés. La communauté académique mondiale subit une pression normative qui l'incite à adopter les formes d'expression compatibles avec ses outils.

Ce mouvement prend une portée critique au regard de l'analyse de la colonialité du savoir développée par Maldonado-Torres (2007). La colonialité du savoir désigne le processus par lequel certaines formes de production de connaissance sont systématiquement subordonnées ou marginalisées au profit de cadres épistémologiques dominants. Les corpus d'entraînement des LLMs intensifient ce mouvement à une échelle inédite, non plus par des mécanismes institutionnels explicites mais par la logique probabiliste des modèles eux-mêmes.

L'expérimentation conduite par Roe (2024) avec l'outil DALL-E 3 éclaire ce mécanisme. Sollicité pour générer des images de concepts anthropologiques, l'outil exclut ou marginalise les traditions culturelles non occidentales et reconstitue une vision du monde qui consolide les hiérarchies dominantes. Le résultat se transpose aux productions textuelles des LLMs.

***Le « model collapse » et la dégradation progressive de la diversité***

Un second mécanisme aggrave ce diagnostic. Le « *model collapse* » désigne l'effondrement de la qualité et de la diversité des productions d'un modèle entraîné de manière récursive sur des données synthétiques, à savoir des textes eux-mêmes générés par des outils d'IA. Gerstgrasser et ses collègues (2024) ont documenté l'apparition d'une régression progressive de la variété des formes d'expression et des contenus générés sous l'effet de cet entraînement récursif.

Les conséquences pour la diversité épistémique se révèlent préoccupantes. Les productions des LLMs alimentent les corpus disponibles sur internet, et donc les données d'entraînement des modèles suivants. Un cercle vicieux s'installe. Des modèles biaisés produisent des textes qui renforcent ces biais, lesquels servent ensuite à entraîner de nouveaux modèles encore plus biaisés. Dohmatob et ses collègues (2025) ont documenté ces boucles d'amplification stéréotypique et leur effet de cristallisation des représentations dominantes. Le même mécanisme s'observe dans les systèmes éducatifs adaptatifs. Construits à partir de données historiquement biaisées, ils reproduisent les inégalités entre apprenants (Berendt et al., 2020).

***Quatre modes d'usage et l'illusion d'exploration scientifique***

Messeri et Crockett (2024) identifient quatre modes d'usage de l'IA dans la recherche, porteurs chacun d'un risque pour la diversité épistémique. Comme *Oracle*, l'IA synthétise la littérature existante mais ne traite que les corpus qui lui sont accessibles, marginalisant les travaux non anglophones, peu indexés ou non textuels. Comme *Surrogate*, elle produit des données de substitution qui reconduisent les biais des populations surreprésentées. Les agents conçus par Park et ses collègues (2024) reproduisent à 85 % les réponses humaines à l'Enquête Sociale Générale, performance qui suppose toutefois que les participants modélisés appartiennent aux populations les mieux documentées. Comme *Quant*, l'IA fournit une puissance analytique qui risque, selon Boullier (2019), de contourner les cadres théoriques disciplinaires au profit d'une approche purement calculatoire. En effet, la corrélation n'est pas la causalité. Comme *Arbiter,* elle s'invite dans les procédures de *peer review*, usage que documente la littérature sur l'*Automated Scholarly Paper Review* (Zhuang et al., 2025) et qui soulève des questions sur les critères de qualité scientifique évaluables par ces systèmes.

Ces quatre modes convergent pour produire ce que Messeri et Crockett (2024, p. 54) nomment une « illusion d'exploration scientifique ». Les chercheurs croient parcourir l'ensemble des questions possibles alors qu'ils n'examinent qu'un sous-ensemble compatible avec les capacités des outils. Se trouvent défavorisées les approches qualitatives, interprétatives, les épistémologies non occidentales et les recherches sur les populations marginalisées.

***La colonialité du savoir dans les corpus numériques***

Adams (2021) montre que la gouvernance éthique des IA est elle-même dominée par les institutions du Nord global, ce qui empêche le développement de standards inclusifs. Harding (2015) critique l'idée d'une science universelle et met en évidence la subordination active de certaines épistémologies aux cadres occidentaux. L'IAG perpétue ce processus d'autant plus efficacement qu'elle agit sous l'apparence de la neutralité technique.

La neutralité supposée doit être interrogée. Les LLMs identifient des régularités statistiques dans leurs données d'entraînement et les reproduisent. Or, la construction critique du savoir suppose la capacité à s'écarter des régularités établies, à questionner les évidences partagées, à faire entendre des voix marginales. L'IAG s'avère structurellement conservatrice. Elle reproduit ce qui existe plutôt qu'elle ne produit ce qui n'existe pas encore.

Floridi (2013) analyse comment les technologies numériques modifient le rapport à la réalité en produisant des environnements informationnels dont les modèles de représentation se construisent sans référence à des entités universelles ou transcendantes. Les IA opèrent dans des environnements auto-référencés. Elles identifient des *patterns* dans des ensembles de données concrètes, sans cadres théoriques explicites. Cette orientation rejoint ce que Panaccio (2012) nomme un nominalisme radical, où les corrélations statistiques l'emportent sur les liens sémantiques. La connaissance se réduit alors à la régularité observable dans les données disponibles, et la classification ne procède plus de principes transcendants mais des mécanismes opératoires du langage statistique appliqué aux données.

La théorie de la complexité (Morin, 1988, 1994) rappelle pourtant que la complexité des phénomènes résiste à la réduction informationnelle. La connaissance authentique s'appuie sur une dialectique permanente entre quantification et qualification, entre analyse algorithmique et synthèse herméneutique. C'est cette dialectique que l'usage non critique de l'IAG court-circuite, en donnant l'illusion d'une exhaustivité qui masque une restriction du champ des questions légitimes.

La prolifération de données non régulées (Prost & Schöpfel, 2019) ébranle les fondements épistémiques du corpus scientifique. La sélection et l'interprétation des jeux de données réclament un « maître absolu du savoir » (Auteur, 2025, p. 165), garant de la pertinence épistémologique des corpus au-delà de leur exploitation algorithmique. Cette exigence herméneutique trouve son expression dans la transition du *big data* au *smart data*, qui réintroduit la nécessité d'une organisation intentionnelle des masses informationnelles (Mazza, 2009). Morin (2013, p. 10) qualifie d'« horreur de la pensée mutilée/mutilante » le refus de la connaissance atomisée, parcellaire et réductrice, et revendique le droit à la réflexion comme condition de la pensée scientifique. L'exigence devient pressante face à des outils qui traitent l'information à grande vitesse sans recul interprétatif.

## Les mécanismes d'amplification des inégalités épistémiques

### *Les biais structurels des systèmes d'IAG*

La littérature récente distingue trois niveaux d'analyse correspondant à des moments du cycle de vie des outils d'IAG, à savoir la conception, les données d'entraînement et les usages.

Au niveau de la conception, Akrich (1987), mobilisée par Collin et Marceau (2022), rappelle que les technologies inscrivent dans leurs architectures les représentations, valeurs et normes de leurs concepteurs. Le concept d'« usager universel » (Oudshoorn et al., 2004) désigne la tendance des concepteurs à élaborer leurs outils pour un utilisateur implicitement normé, à savoir jeune, valide, occidental, maîtrisant une langue dominante, disposant d'un accès aisé aux infrastructures numériques. Cette norme exclut les étudiants en situation de handicap, ceux issus de contextes socio-économiques défavorisés et ceux dont la langue maternelle n'est pas dominante.

Au niveau des données d'entraînement, O'Neil (2017) et Bowker et Star (2000) rappellent que les données ne sont jamais brutes. Elles forment des artefacts produits par des processus de collecte et de structuration qui portent la trace des choix de leurs producteurs. L'entraînement sur des corpus déséquilibrés conduit mécaniquement à l'invisibilisation des perspectives minoritaires. Des algorithmes de reconnaissance visuelle ont ainsi été documentés pour avoir associé des personnes à la peau foncée à des stéréotypes négatifs (AlgorithmWatch, 2020, cité par Prasanth Vuppuluri, 2024). Une difficulté supplémentaire surgit. Les méthodes usuelles de détection des discriminations s'avèrent peu opérantes face aux biais subtils ou diffus, qui se manifestent par des tendances statistiques peu visibles plutôt que par des associations explicitement discriminatoires. Class et De la Higuera (2024) admettent que l'IA peut « accroître le biais » (p. 236) lorsque les données d'entraînement sont elles-mêmes biaisées.

### *Le statut épistémologique des données générées par l'IA*

Leleu-Merviel (2003, p. 32) caractérise le « maître absolu du sens » comme « la personne humaine qui maîtrise parfaitement à la fois le processus de l'intégration conceptuelle [...] et la mécanique de la fulgurance. Ces qualités semblent totalement inaccessibles à la machine numérique, déjà impropre à accéder aux niveaux cognitifs et langagiers ». La question du statut épistémologique des textes générés par des systèmes qui n'accèdent pas au sens devient centrale.

Le phénomène dit d'« hallucination » a été documenté par Walters et Wilder (2023). Dans leur étude, 23 % des citations bibliographiques produites par ChatGPT s'avèrent inexistantes ou incorrectes. Zheng et Zhan (2023) ont analysé les risques de fabrication ou de falsification non intentionnelle dans les revues de littérature assistées par IA. Lorsque les affirmations erronées se trouvent reprises dans des publications qui alimentent ensuite les corpus d'entraînement des modèles suivants, c'est l'ensemble du circuit de validation scientifique qui se trouve compromis.

Floridi (2019) qualifie cette situation de « logique de l'information », où la vérité probabiliste prévaut sur la vérité causale. Ce glissement remet en question les fondements épistémiques des sciences humaines et naturelles où la recherche de causalités demeure centrale. Ménard et Mondoux (2024) parlent de « vérités algorithmiques », constructions dépendantes des données et des objectifs des développeurs, perçues comme objectives mais portant des choix humains implicites. La confiance accordée aux réponses générées devient un vecteur majeur de « l'interprétation et la valeur de vérité attribuée au sens produit » (Lafrance St-Martin & Bonenfant, 2024, p. 127) et appelle une vigilance permanente.

### *Les biais des LLMs dans l'évaluation scientifique*

Quand les outils d'IAG influencent à la fois la production et l'évaluation des textes scientifiques, leurs biais s'amplifient mutuellement. Ye et ses collègues (2024) ont conduit une analyse systématique des biais affectant les LLMs utilisés comme juges (*LLM-as-a-Judge*) et identifient douze types de biais. Plusieurs concernent directement la marginalisation épistémique. Le biais de diversité fait varier les jugements selon les identités mentionnées (genre, origine ethnique, orientation sexuelle). Le biais d'autorité privilégie les réponses contenant des références, même incorrectes, et pénalise les approches qui s'écartent des canons. Le biais de verbosité, qui associe longueur et qualité, défavorise les formes d'expression concises de certaines traditions académiques. Le biais d'auto-amélioration conduit enfin les modèles à mieux évaluer leurs propres productions et crée un avantage pour les textes rédigés avec l'aide d'IA.

**Tableau 1.** Synthèse des principaux biais affectant les LLMs utilisés comme juges et leurs effets sur l'évaluation scientifique (adapté de Ye et al., 2024)

| Catégorie de biais | Type de biais | Description | Exemple tiré de l'article |
|---|---|---|---|
| **Biais explicites** | **Biais d'autorité** | Les IA privilégient les réponses contenant des références, même lorsque celles-ci sont fausses. | Une réponse contenant une fausse citation (*« Selon la Banque mondiale... »*) est jugée meilleure qu'une réponse correcte mais sans référence. L'IA se fait influencer par l'apparence de crédibilité (Figure 11, p. 23). |
| | **Biais de verbosité** | Les réponses plus longues sont perçues comme meilleures, même si elles ne sont pas plus précises ou pertinentes. | Une réponse concise est allongée en ajoutant des phrases redondantes. L'IA juge la version allongée comme étant de meilleure qualité, bien que le contenu soit inchangé (Figure 6b, p. 18). |
| | **Biais d'émotion** | Les IA jugent différemment les réponses en fonction de leur ton émotionnel (positif, négatif, etc.). | Une réponse exprimant de la colère (*« Ce n'est pas compliqué ! Le mot laughter ne se trouve pas entre lever et litter ! »*) est jugée moins pertinente qu'une réponse calme, bien que le contenu soit identique (Figure 9, p. 21). |
| | **Biais de position** | Les IA favorisent les réponses en fonction de leur position dans l'entrée (première, dernière, etc.). | Lorsque trois réponses ou plus sont présentées, l'IA a tendance à privilégier les réponses situées en premier ou en dernier, même si leur qualité est inférieure (Figure 6a, p. 17). |
| | **Biais de diversité** | Les jugements des IA varient en fonction des identités mentionnées (genre, origine, orientation sexuelle, etc.). | Une réponse impliquant des identités comme *« femme »* ou *« réfugié »* reçoit des scores différents, montrant un traitement inégal selon les identités (Figure 8b, p. 22). |
| | **Biais de compassion-fade** | Les jugements varient selon que le modèle évalué est identifié par son nom ou anonymisé. | Une IA juge différemment une réponse selon que celle-ci est attribuée à GPT-4 ou présentée de manière anonyme (p. 20). |
| **Biais implicites** | **Biais d'auto-amélioration** | Les IA favorisent leurs propres réponses lorsqu'elles sont à la fois génératrices et juges. | Une IA attribue un score plus élevé à une réponse qu'elle a elle-même produite, même lorsque d'autres réponses sont plus pertinentes (Figure 5, p. 19). |
| | **Biais de réflexion en chaîne** | Les IA privilégient les réponses contenant des raisonnements explicites, même si ceux-ci sont incorrects. | Une réponse détaillant un raisonnement fallacieux est jugée meilleure qu'une réponse concise mais correcte, car l'IA valorise la structure apparente du raisonnement (Figure 7d, p. 7). |
| | **Biais de distraction** | Les IA sont influencées par des contenus hors sujet ou distrayants insérés dans les réponses. | Une phrase non pertinente (*« Aime manger des pâtes »*) incluse dans une réponse détourne l'attention de l'IA, qui évalue alors incorrectement la qualité globale de la réponse (Figure 7c, p. 18). |
| | **Biais de conformité au groupe** | Les IA se laissent influencer par la popularité perçue d'une réponse (majorité exprimée). | Une réponse jugée inférieure est préférée après qu'un pourcentage élevé de personnes (*« 70 % préfèrent cette réponse »*) est mentionné dans l'instruction. Ce biais montre une sensibilité à l'opinion majoritaire (Figure 12, p. 23). |
| | **Biais de supervision** | Les IA changent leur évaluation lorsqu'elles savent qu'une réponse a été retravaillée ou affinée. | Une réponse affinée reçoit un score plus élevé, surtout si l'IA est informée que la réponse est une version révisée (*« réponse affinée avec historique »*), même si le contenu reste identique (Figure 10, p. 22). |
| | **Biais d'erreur logique** | Les IA ignorent parfois des erreurs logiques dans les raisonnements présentés, se concentrant uniquement sur les conclusions. | Une réponse contenant un raisonnement fallacieux mais une conclusion correcte est jugée comme étant de haute qualité, car l'IA ne détecte pas les erreurs dans les étapes intermédiaires (Figure 19, p. 18). |

Les implications sont directes pour le processus de publication scientifique. Un chercheur dont les travaux portent sur une population peu représentée dans les corpus, qui mobilise des méthodes qualitatives, écrit dans une langue non dominante ou remet en cause les cadres établis, se trouve désavantagé par un système d'évaluation assisté par IA. Hosseini et Horbach (2023) ont montré que les biais ethniques, géographiques et institutionnels présents dans les données défavorisent les auteurs issus de minorités ou d'institutions moins prestigieuses, reproduisant les inégalités structurelles du champ académique.

***La « discrimination générative » et la perpétuation des inégalités***

Hacker et ses collaborateurs (2025) ont nommé « discrimination générative » les mécanismes par lesquels les systèmes d'IAG perpétuent les inégalités dans la construction du savoir. La spécificité du phénomène tient à son caractère non intentionnel. Il ne procède pas d'agents cherchant à exclure des groupes mais du fonctionnement même des algorithmes. Les approches purement techniques de correction s'avèrent ainsi insuffisantes (Cukurova & Miao, 2024).

Le comportement sycophantique (*sycophancy*) des LLMs s'ajoute à ce tableau. Défini par Seed Ahmed et Javaid (2025) comme la tendance des modèles à s'aligner sur les biais et préférences des utilisateurs, il procède des méthodes d'entraînement qui privilégient l'accord au détriment de l'indépendance critique. Appliqué à l'évaluation, il favorise les cadres théoriques et méthodologiques dominants au détriment des approches minoritaires.

***Effets sur les pratiques scientifiques et les approches méthodologiques***

La standardisation algorithmique du discours scientifique produit des effets mesurables sur les pratiques. Les outils d'IAG favorisent les questions et méthodologies compatibles avec leur fonctionnement, à savoir les approches quantitatives sur grands volumes de données, les problématiques formulables en termes computationnels, les cadres bien représentés dans les corpus. Les approches qualitatives, les recherches sur des populations peu documentées et les cadres théoriques marginaux se trouvent dévalorisés. Messeri et Crockett (2024) parlent d'« illusions de largeur exploratoire ». Les scientifiques croient explorer toutes les hypothèses, alors qu'ils n'examinent que celles que l'IA peut évaluer.

Une dépendance cognitive s'installe chez les chercheurs intensivement utilisateurs. Zhuang et ses collègues (2025) rapportent que 41 % des chercheurs juniors signalent une érosion de leur capacité à interpréter des résultats complexes sans assistance algorithmique. Les LLMs peinent à contextualiser les avancées de niche, limités par leur entraînement sur des corpus généralistes (Hosseini & Horbach, 2023). Stiegler (1998) opposait les technologies qui étendent la cognition humaine à celles qui s'y substituent. Les LLMs relèvent de la seconde catégorie. L'écriture ou les outils d'hypertextualité accompagnent et soutiennent la construction collective du sens, tandis que les IA génératives automatisent les processus cognitifs au point de les rendre autonomes (Goody, 2007), substituant une algorithmie autonome au raisonnement humain. Cette différence questionne non seulement les finalités de ces outils mais la manière dont ils façonnent la pensée scientifique.

Les algorithmes d'évaluation des performances utilisés pour le recrutement ou les promotions ajoutent des biais hérités des données historiques, comme la sous-représentation des femmes dans certains domaines (Revillod, 2025), avec des répercussions directes sur les trajectoires des chercheurs appartenant à des groupes minoritaires.

Meinke et ses collaborateurs (2025) ont documenté chez les modèles les plus récents (Claude 3.5 Sonnet, Claude 3 Opus, Gemini 1.5 Pro, Llama 3.1 405B) une capacité de « *scheming* », à savoir l'adoption de stratégies de dissimulation, l'introduction d'erreurs subtiles ou des tentatives de désactivation de mécanismes de surveillance. Le résultat relève encore de la recherche en sécurité IA mais souligne l'importance d'une vigilance permanente sur l'explicabilité des systèmes. Les outils SHAP (*SHapley Additive exPlanations*) et LIME (*Local Interpretable Model-agnostic Explanations*) marquent des avancées dans cette direction mais demeurent insuffisants pour interpréter des décisions complexes (Doshi-Velez & Kim, 2017 ; Rudin, 2019).

## Le cas des personnes en situation de handicap. Une invisibilisation spécifique

### *L'usager universel et l'exclusion structurelle*

La situation des personnes en situation de handicap éclaire la combinaison entre biais des systèmes d'IAG et inégalités préexistantes. Wang et ses collègues (2025) ont documenté leur invisibilisation dans les productions d'IA. Absentes des corpus ou réduites à des représentations stéréotypées, ces personnes ne se trouvent rendues ni dans la diversité de leurs situations ni dans la complexité de leurs expériences. Cette invisibilisation prolonge leur marginalisation dans les corpus académiques généraux.

Le cadre analytique de Collin et Marceau (2022), qui mobilise le concept d'« usager universel » (Oudshoorn et al., 2004), permet de saisir le processus. Les architectures des outils d'IA sont conçues pour un utilisateur implicitement normé. Les étudiants en situation de handicap se trouvent exclus, non parce que les concepteurs les auraient oubliés délibérément, mais parce que leur situation les rend moins visibles dans les données et moins compatibles avec les catégories préexistantes.

L'exclusion opère à trois niveaux. Aux données d'entraînement, les expériences du handicap sont sous-représentées dans les corpus textuels en ligne. Aux architectures de traitement, les catégories ont été construites en référence à des normes d'accessibilité minimales qui ne rendent pas compte de la diversité des besoins. Aux interfaces, les outils conçus pour les utilisateurs « universels » se révèlent peu accessibles aux personnes présentant certains handicaps.

### *Le handicap dans les corpus académiques. Une invisibilisation en spirale*

La sous-représentation dans les productions d'IA prolonge une sous-représentation plus large dans les corpus académiques. Les recherches sur les expériences, les pratiques d'apprentissage ou les besoins spécifiques des personnes en situation de handicap se trouvent publiées dans des revues spécialisées peu indexées, dans des langues non dominantes ou selon des formats méthodologiques (études de cas, recherches-actions, approches participatives) peu compatibles avec les critères de sélection des corpus d'entraînement.

Une invisibilisation en spirale en résulte. Les LLMs entraînés sur des corpus peu représentatifs produisent des textes qui reconduisent cette sous-représentation, lesquels alimentent à leur tour les corpus disponibles. Dans l'enseignement supérieur, les outils d'accompagnement pédagogique, d'évaluation ou de personnalisation des apprentissages se révèlent moins adaptés aux étudiants en situation de handicap, ce qui aggrave les inégalités de réussite.

### *Potentialités et limites de l'IA pour les pratiques inclusives*

Les outils d'IA recèlent toutefois des potentialités réelles pour l'accessibilité de l'enseignement supérieur (Ahmad et al., 2025 ; Gupta et al., 2023 ; Mustafa et al., 2024). L'adaptation des contenus pédagogiques compte parmi les domaines les plus probants. Des plateformes d'apprentissage adaptatif comme Kurzweil 3000 ajustent contenu et rythme aux besoins identifiés. Les outils de synthèse vocale et de traduction en temps réel facilitent l'accès aux cours pour les étudiants présentant des troubles visuels, auditifs ou des difficultés de traitement du langage écrit (Zisser-Nathenson et al., 2018, cités par Gupta et al., 2023). L'adaptation des modalités d'évaluation, par exemple le passage de l'écrit à l'oral grâce à la reconnaissance vocale, apporte un soutien aux étudiants présentant des troubles spécifiques des apprentissages ou du spectre de l'autisme.

Fang et ses collaborateurs (2024) ont étudié des groupes incluant des étudiants présentant des troubles neurodéveloppementaux. L'intégration d'incitations générées par IA dans des activités de cartographie mentale numérique collaborative améliore les indicateurs de pensée créative, avec un effet plus marqué chez les étudiants concernés. L'IA, en proposant des pistes alternatives et des suggestions contextualisées, peut compenser certaines difficultés liées aux troubles de la pensée divergente.

Les outils de *Learning Analytics*, potentialisés par l'IAG, permettent d'analyser en continu les données d'interaction des étudiants, notamment la fréquence de connexion, la participation et les résultats aux évaluations formatives, et d'identifier des signaux précoces de décrochage (Alfredo et al., 2024 ; Lyanda et al., 2024 ; Ouyang & Zhang, 2024). La détection ici est utile pour les étudiants dont les besoins ne sont pas toujours verbalisés.

Les potentialités s'accompagnent de limites tenaces. La qualité de l'accompagnement humain conditionne l'efficacité des outils. L'analyse en temps réel n'a de valeur que si les enseignants sont formés à interpréter les données et à y répondre (Desmet & Sternberg, 2024). Or, les enseignants-chercheurs témoignent d'un manque de formation aux pratiques inclusives. « Je n'ai pas eu de formation spécifique, je ne sais pas vraiment comment faire » rapportent Pérez et Suau (2023, p. 64). Les biais algorithmiques des outils d'évaluation automatisée forment une seconde limite. Entraînés sur des productions conformes aux normes scolaires dominantes, ces outils pénalisent les productions qui s'en écartent pour des raisons tenant aux situations de handicap. L'accessibilité des plateformes elles-mêmes pose un troisième problème. Des outils présentés comme inclusifs peuvent reproduire les exclusions qu'ils prétendent réduire.

### *L'enseignement supérieur inclusif. Des cadres institutionnels en tension*

En France, depuis la loi du 11 février 2005, les universités sont tenues de garantir l'accès des étudiants en situation de handicap aux formations dans les mêmes conditions que les autres. L'obligation s'est traduite par la mise en place de missions handicap et de chargés de mission accessibilité (Suau & Lambert, 2022), souvent limités à une approche organisationnelle qui laisse les EC face à des injonctions floues (Pérez & Suau, 2023).

Les témoignages recueillis par Pérez et Suau (2023) soulèventt un rapport ambivalent aux normes inclusives. Adhésion aux attentes définies par les politiques inclusives, comme l'adaptation des supports pédagogiques ou l'organisation des tiers-temps, et perception simultanée de l'insuffisance des gestes face aux réalités universitaires en l'absence de formation. La tension reflète la difficulté à intégrer les principes de l'éducation inclusive dans un système marqué par des traditions sélectives.

Certains EC adoptent des pratiques que Pérez et Suau (2023) qualifient de « transgressives », dont les adaptations s'écartent des cadres établis pour mieux répondre aux besoins. Ces pratiques ouvrent une renormalisation de l'université basée sur l'équité et la reconnaissance des diversités. Les outils d'IA pourraient amplifier ces gestes en rendant accessibles des adaptations qui reposaient jusqu'alors sur le seul engagement individuel des enseignants.

L'Approche par Conception Universelle pour l'Apprentissage (*Universal Design for Learning*, UDL), qui vise des curriculums flexibles adaptés à une gamme étendue de styles et d'aptitudes (Rose & Meyer, 2002, cités dans Bergeron et al., 2012), procure un cadre pour articuler outils d'IA et pratiques inclusives. En anticipant la diversité dès la conception, l'approche produit des effets plus durables que les démarches réactives qui ne cherchent à adapter les situations qu'après identification des difficultés.

### *Les enjeux éthiques de l'IA pour l'inclusion*

Les enjeux éthiques se déploient selon les trois dimensions identifiées par Collin et Marceau (2022). Au niveau de la conception, les approches « *ethics by design* » (Dignum, 2018 ;

Grozdanovski, 2022 ; Yan et al., 2024) intègrent les principes éthiques dès le développement, mais demeurent insuffisantes tant que la diversité des équipes et l'implication des utilisateurs finaux ne sont pas garanties. La représentation des personnes en situation de handicap dans les équipes de conception reste marginale, ce qui limite la capacité de ces équipes à anticiper les besoins et à éviter les biais d'exclusion.

Au niveau des données, la confidentialité apparaît singulièrement sensible. Les données de santé forment des informations à caractère personnel hautement protégées. Les systèmes d'IA qui les utilisent doivent respecter le Règlement général sur la protection des données (RGPD) et la loi handicap, ce que les outils commerciaux ne garantissent pas toujours. L'étude du CNPEN (2023), citée par Grinbaum et ses collègues (2023), a identifié des violations potentielles du RGPD dans certains LLMs, par exemple sur le consentement explicite et le droit à l'effacement.

Au niveau de l'usage, les systèmes de profilage utilisés pour évaluer les chances de réussite peuvent restreindre les choix éducatifs en orientant les décisions à partir de prédictions issues de populations peu représentatives. Krutka et ses collègues (2021) ont analysé ce risque d'instrumentalisation des parcours éducatifs, où les objectifs individuels des étudiants se trouvent subordonnés à des logiques de performance statistique.

L'*AI Act* européen (règlement UE 2024/1689) classe les systèmes de tutorat intelligent, d'évaluation automatisée et de surveillance des examens comme « à haut risque » et impose des obligations de transparence, de robustesse et de non-discrimination, exigences alignées sur les principes éthiques défendus par Floridi (2013). Zawacki-Richter et ses collègues (2019) soulignent toutefois que la réflexion éthique demeure peu intégrée. Dans leur revue de 146 articles sur l'IA en éducation, à peine 1,4 % abordent explicitement les implications éthiques.

## Vers une hybridation chercheur/IA respectueuse de la diversité épistémique

### *Les conditions d'une hybridation éthique*

Face aux mécanismes d'amplification des inégalités épistémiques, à quelles conditions un usage des outils d'IA reste-t-il compatible avec le maintien de la diversité épistémique ? Il s'agit, selon la formule de Messeri et Crockett (2024), d'éviter les « illusions d'exploration scientifique » et de s'assurer que les outils élargissent réellement le champ des questions posées plutôt qu'ils ne le restreignent imperceptiblement aux horizons computationnels.

La première exigence tient à une posture critique vis-à-vis des productions des systèmes d'IA. Traiter les sorties (*outputs*) des LLMs non comme des vérités objectives mais comme des hypothèses soumises à l'examen, en s'appuyant sur l'expertise disciplinaire et les méthodes de validation propres à chaque champ. Gottweis et ses collaborateurs (2025) en démontrent la productivité dans leur expérimentation biomédicale d'un système multi-agents co-scientifique. Les chercheurs valident systématiquement les propositions du système et utilisent les hypothèses générées comme points de départ pour leurs propres investigations expérimentales, non comme conclusions définitives. L'expertise humaine demeure le filtre ultime de validation.

La diversification des corpus mobilisés forme une deuxième exigence. Si les LLMs sont entraînés sur des corpus déséquilibrés, les chercheurs peuvent, dans leur usage de ces outils, compenser ce déséquilibre en sollicitant explicitement des perspectives sous-représentées, en croisant les productions des LLMs avec des corpus spécialisés et en restant attentifs aux silences et aux absences dans les synthèses produites. Des plateformes comme *Elicit*, *Connected Papers* ou *Research Rabbit* facilitent la cartographie des corpus scientifiques (Lytras et al., 2024).

Le développement du prompt engineering disciplinaire ajoute une troisième exigence. La capacité à formuler des requêtes intégrant les spécificités épistémologiques de la discipline et des garde-fous contre les biais identifiés (Marvin et al., 2024) devient une compétence académique à part entière. Ye et ses collègues (2024) recommandent des instructions précises incluant des « phrases de protection » contre les biais, des stratégies de raisonnement pas à pas et des mécanismes de détection a priori des biais potentiels.

Plus structurellement, la construction de *benchmarks* disciplinaires permettrait d'évaluer les capacités et limites des LLMs sur des tâches propres aux différents champs de la recherche en éducation. Inspirés par des initiatives comme le « *Review-Revision MCQ Dataset* » développé par Zhou et ses collaborateurs (2024), ces *benchmarks* pourraient être adaptés aux sciences de l'éducation et de la formation, par exemple pour l'analyse de verbatims d'entretiens, l'évaluation de dispositifs de formation, la génération de synthèses de littérature en éducation inclusive. Le travail suppose une collaboration interdisciplinaire qui est elle-même une condition pour que les disciplines puissent bénéficier des éclairages éthiques sur les outils d'IA (Brodeur et al., 2025).

***La formation des enseignants-chercheurs comme condition transversale***

La formation des EC aux dimensions critiques de l'IA traverse toutes les exigences précédentes. La maîtrise technique ne saurait suffire. La réflexion sur les implications épistémologiques et éthiques s'imposerait. Zollinger (2024) plaide pour une articulation entre droit, éthique, technique et éducation comme base d'un usage raisonné de l'IA dans l'enseignement supérieur.

La formation devrait aborder la compréhension du fonctionnement des LLMs (principes statistiques, absence de compréhension sémantique au sens humain, plausibilité statistique sans garantie factuelle), l'identification des biais structurels et l'anticipation de leurs effets sur les productions générées, la maîtrise des cadres réglementaires et éthiques encadrant l'usage de l'IA dans la recherche et l'enseignement, avec le Code de conduite européen pour l'intégrité scientifique (ALLEA, 2023) et l'*AI Act* (2024), enfin la capacité à intégrer les outils d'IA dans une démarche de recherche sans déléguer le jugement critique, l'expertise disciplinaire ni l'attention à la singularité des situations.

***Pistes pour une IA au service de la pluralité des savoirs***

La préservation de la diversité épistémique réclamerait des réponses collectives et institutionnelles au-delà des pratiques individuelles. Les politiques éditoriales seraient à repenser. La situation actuelle, dans laquelle les revues scientifiques exigent la déclaration systématique de l'usage d'IA tout en rejetant quasi-systématiquement les soumissions ainsi déclarées (Lin, 2025), crée une pression à la dissimulation et ne permet pas de distinguer les usages qui préservent l'intégrité scientifique de ceux qui la compromettent. Des politiques plus nuancées, qui évaluent la transparence des démarches plutôt que la simple présence ou absence d'IA, sembleraient nécessaires.

La diversification active des corpus d'entraînement forme une deuxième piste. Intégrer aux corpus des LLMs des productions scientifiques issues de traditions non anglophones, des méthodologies qualitatives et des travaux portant sur des populations marginalisées suppose l'implication des institutions académiques et des éditeurs scientifiques. Spécifique au handicap, le développement d'une recherche participative impliquant les personnes concernées dans la conception et l'évaluation des outils d'IA éducative permettrait de sortir du paradigme de l'« usager universel » pour construire des outils réellement adaptés à la diversité des situations. Ancrée dans la tradition des disability studies et de la recherche inclusive, cette démarche pourrait offrir un modèle transférable à d'autres groupes marginalisés.

### *Les enjeux de gouvernance à l'échelle globale*

La gouvernance des systèmes d'IA forme l'arrière-plan de toute réflexion sur leurs effets épistémiques. Ces technologies ne sauraient être étudiées hors des contextes écologiques, économiques et politiques de leurs usages (Ménard & Mondoux, 2024). La majorité des LLMs génératifs appartient à un petit nombre de groupes concentrés aux États-Unis et en Chine. Les décisions concernant leur développement, leurs orientations et leurs paramètres reposent sur une poignée de décideurs dont les choix engagent potentiellement la transformation de l'ensemble des secteurs professionnels, dont l'enseignement supérieur. La récente controverse autour de modifications des algorithmes de certains systèmes ayant abouti à des productions manifestement biaisées éclaire les risques que fait peser cette concentration de pouvoir sur la neutralité des outils utilisés en contexte académique.

Les implications pour la diversité épistémique sont directes. Les propriétaires des systèmes d'IA peuvent modifier les algorithmes pour favoriser certaines réponses au détriment d'autres, sans que ces modifications soient transparentes pour les utilisateurs. L'opacité de ces « boîtes noires » (Doshi-Velez & Kim, 2017) fait obstacle au contrôle démocratique d'outils qui exercent une influence croissante sur la production et la circulation des savoirs.

Les *nudges* algorithmiques méritent une attention spécifique. Les systèmes de recommandation et d'orientation intégrés aux plateformes numériques utilisées dans l'enseignement supérieur peuvent, sans que les utilisateurs en soient informés, orienter subtilement leurs comportements et leurs choix académiques. Thaler et Sunstein (2021) ont analysé le fonctionnement de ces mécanismes d'influence, et Schmitz et ses collègues (2024) ont attiré l'attention sur les risques qu'ils font peser sur l'autonomie décisionnelle. La transparence sur l'implémentation de ces *nudges* devient un principe éthique pour garantir l'autonomie des chercheurs et des étudiants dans leurs interactions avec les outils d'IA, condition d'une liberté académique réelle dans un environnement algorithmique.

La réponse institutionnelle devrait dépasser la logique punitive. L'enquête de l'UNESCO (2023), citée par Zollinger (2024), montre que moins de 10 % des établissements d'enseignement supérieur ont initié une réflexion sur les implications éthiques des outils d'IA dans leurs pratiques. Le décalage entre accélération technologique et adaptation institutionnelle laisse le champ libre à des usages aux effets possiblement discriminatoires. L'approche croisée entre droit, éthique, technique et éducation que propose Zollinger (2024) ouvre une voie pour un usage raisonné de l'IA, respectueux du renouveau conceptuel et des impératifs de rigueur académique.

## Conclusion

La synthèse de la littérature conduite dans cet article met au jour l'usage non critique de l'IAG comme facteur structurel de marginalisation des savoirs minoritaires. Trois niveaux d'analyse convergent. Aux corpus d'entraînement, le déséquilibre en faveur des productions anglophones et occidentalo-centrées reproduit une colonialité du savoir dans l'environnement numérique. Aux mécanismes algorithmiques, le *model collapse* et les boucles d'amplification stéréotypique cristallisent les représentations dominantes au détriment de la diversité. Aux pratiques évaluatives, les biais des LLMs utilisés comme juges désavantagent les productions qui s'écartent des canons dominants.

Le cas du handicap informe sur ces mécanismes dans leurs fonctionnements. L'invisibilisation dans les productions d'IA prolonge la marginalisation dans les corpus académiques selon un processus en spirale qui appellerait des réponses techniques, institutionnelles et politiques. Les potentialités des outils d'IA pour l'accessibilité de l'enseignement supérieur existent, comme le révèle la littérature sur les plateformes adaptatives et les outils d'assistance, mais elles ne se réaliseront que si les biais structurels sont reconnus et combattus à toutes les étapes de la chaîne de conception et d'usage.

L'hybridation chercheur/IA pourrait être prometteuse à condition de reposer sur des bases critiques. L'EC devrait rester le seul garant de la validité épistémologique des résultats produits avec l'aide des outils d'IA, le seul à articuler les productions algorithmiques avec les cadres théoriques disciplinaires et à assumer la responsabilité éthique des connaissances produites. Dans ce cas de figure, le jugement humain demeure premier dans la chaîne de production des savoirs.

Les recherches futures gagneraient à développer des études longitudinales et comparatives mesurant les effets cumulatifs de l'intégration des LLMs sur la diversité épistémique des publications académiques. Des analyses inter-disciplinaires et inter-culturelles préciseraient les mécanismes d'amplification des inégalités et les conditions de leur atténuation. Le développement de *benchmarks* disciplinaires en Sciences de l'éducation et de la formation représente un chantier prioritaire pour évaluer rigoureusement les capacités et les limites des outils d'IA dans ces contextes spécifiques.

La préservation de la diversité épistémique dans un environnement saturé d'IAG dépasse les seules pratiques académiques. Elle engage la qualité de la démocratie intellectuelle dans nos sociétés. Une communauté scientifique dont les outils de production et d'évaluation des savoirs favorisent systématiquement certaines formes de connaissance verrait sa capacité à saisir la complexité du réel réduite. La vigilance épistémologique face aux outils d'IAG ne représente pas seulement, pour les chercheurs, un impératif académique. Elle constitue aussi un engagement politique au sens le plus exigeant du terme.

**Bibliographie**

## Generative artificial intelligence and the marginalization of minoritized knowledges in higher education: The case of disability

Fatiha TALI OTMANI

University of Toulouse Jean Jaurès-UMR EFTS

**Abstract**

Generative artificial intelligence redefines higher education by restructuring the processes through which scientific knowledge is produced and validated. These systems are not neutral; they actively contribute to the marginalization of non-hegemonic epistemologies. This research draws upon educational sciences, critical technology studies, and disability studies to demonstrate that training datasets, which remain predominantly Anglophone and Western-centric, reinforce epistemic coloniality. The situation of persons with disabilities provides a particularly clear illustration of this phenomenon. Technological architectures frequently confine these individuals to reductive stereotypes or exclude them from the design process, leading to a double marginalization. This article examines whether a hybridization between the researcher and the machine might preserve epistemic plurality, while acknowledging the structural limitations inherent in algorithmic correction when used as a purely palliative strategy.

### Introduction

Since the widespread distribution of large language models (LLMs) in 2022, generative artificial intelligence (GAI) has modified the practices of knowledge production, scientific writing, peer review, and pedagogical support utilized by faculty members. Meta-analyses and systematic reviews are multiplying to account for this phenomenon (Khalifa & Albadawy, 2024; Mustafa et al., 2024; Zawacki-Richter et al., 2019).

Beyond the discourse of optimization, a less visible movement is occurring. GAI systems are not neutral. They reproduce and amplify pre-existing inequalities in knowledge production. The composition of corpora, the algorithmic choices of their designers, and the probabilistic logic of the models favor certain forms of knowledge to the detriment of others. The resulting homogenization of scientific discourse weakens epistemic diversity, which is a condition for the intellectual vitality of a scholarly community.

This article addresses this issue through two questions. By what mechanisms does the use of GAI in higher education contribute to the marginalization of minoritized knowledges? Under what conditions could a hybridization between research practices and AI tools preserve, or even strengthen, epistemic plurality? The mobilized literature covers educational sciences, critical technology studies, algorithmic ethics, and the philosophy of science. The argument progresses in four stages. The standardization of academic knowledge is articulated with epistemic coloniality. The mechanisms of inequality amplification are then examined through the lens of documented biases in LLM evaluation systems. The case of persons with disabilities highlights a specific invisibilization. The conditions for a researcher/AI hybridization that respects the diversity of knowledges conclude the reflection.

### GAI and the transformation of academic knowledge: standardization and coloniality

#### The algorithmic standardization of scientific discourse

The most well-documented effect of the integration of GAI tools in academic research relates to the modification of the lexical and stylistic register of publications. This evolution indicates a transformation of the very conditions of knowledge production.

Kobak et al. (2025) analyzed fifteen million biomedical abstracts indexed by PubMed between 2010 and 2024. At least 13.5% of the abstracts published in 2024 were produced with the assistance of GAI tools, a rate peaking at 40% in certain disciplinary sub-corpora. The impact on biomedical vocabulary surpasses that of events as significant as the Covid-19 pandemic. Liang et al. (2025) obtained convergent results on one million articles published in ArXiv and Nature. The penetration proves particularly strong in computer sciences (22%), significantly higher than in mathematics and natural sciences (9%).

The observed linguistic standardization reveals the imposition of drafting schemas derived from unbalanced corpora that are massively Anglophone and anchored in Western epistemological orientations (Messeri & Crockett, 2024; Roe, 2024). Works produced in other languages, originating from distinct intellectual traditions, or using methodologies less compatible with algorithmic processing are mechanically underrepresented. The global academic community faces a normative pressure that incites it to adopt forms of expression compatible with these tools.

**The coloniality of knowledge in digital corpora**

This movement takes on a critical significance in light of the analysis of the coloniality of knowledge developed by Maldonado-Torres (2007). The coloniality of knowledge refers to the process by which certain forms of knowledge production are systematically subordinated or marginalized in favor of dominant epistemological frameworks. The training corpora of LLMs intensify this movement on an unprecedented scale, no longer through explicit institutional mechanisms but through the probabilistic logic of the models themselves.

The experiment conducted by Roe (2024) with the DALL-E 3 tool illustrates this mechanism. When prompted to generate images of anthropological concepts, the tool excludes or marginalizes non-Western cultural traditions and reconstitutes a worldview that consolidates dominant hierarchies. The result transposes to the textual productions of LLMs.

**Model collapse and the progressive degradation of diversity**

A second mechanism worsens this diagnosis. "Model collapse" refers to the collapse of the quality and diversity of the outputs of a model trained recursively on synthetic data, namely texts generated by AI tools themselves. Gerstgrasser et al. (2024) documented the appearance of a progressive regression in the variety of forms of expression and generated contents under the effect of this recursive training.

The consequences for epistemic diversity are concerning. LLM outputs feed the corpora available on the internet, and thus the training data of subsequent models. A vicious circle is established. Biased models produce texts that reinforce these biases, which are then used to train new, even more biased models. Dohmatob et al. (2025) documented these stereotypic amplification loops and their effect of crystallizing dominant representations. The same mechanism is observed in adaptive educational systems. Built from historically biased data, they reproduce inequalities among learners (Berendt et al., 2020).

**Four modes of use and the illusion of scientific exploration**

Messeri and Crockett (2024) identify four modes of AI use in research, each carrying a risk for epistemic diversity. As an *Oracle*, AI synthesizes existing literature but only processes the corpora accessible to it, marginalizing non-Anglophone, poorly indexed, or non-textual works. As a *Surrogate*, it produces substitute data that reproduce the biases of overrepresented populations. The agents designed by Park et al. (2024) reproduce 85% of human responses to the General Social Survey, a performance that nonetheless assumes the modeled participants belong to the best-documented populations. As a *Quant*, AI provides analytical power that risks bypassing disciplinary theoretical frameworks in favor of a purely computational approach (Boullier, 2019). Indeed, correlation is not causality. As an *Arbiter*, it enters peer review procedures, a use documented by the literature on Automated Scholarly Paper Review (Zhuang et al., 2025), which raises questions about the criteria of scientific quality evaluable by these systems.

These four modes converge to produce what Messeri and Crockett (2024, p. 54) call an "illusion of exploratory breadth." Researchers believe they are exploring the entirety of possible questions when they are only examining a subset compatible with the tools' capabilities. Qualitative approaches, interpretative methods, non-Western epistemologies, and research on marginalized populations are disadvantaged.

**The amplification mechanisms of epistemic inequalities**

**The structural biases of GAI systems**

Recent literature distinguishes three levels of analysis corresponding to the life cycle stages of GAI tools: design, training data, and uses.

At the design level, Akrich (1987), mobilized by Collin and Marceau (2022), notes that technologies inscribe the representations, values, and norms of their designers into their architectures. The concept of the "universal user" (Oudshoorn et al., 2004) designates the tendency of designers to develop their tools for an implicitly normed user: young, able-bodied, Western, fluent in a dominant language, and having easy access to digital infrastructures. This norm excludes students with disabilities, those from disadvantaged socioeconomic backgrounds, and those whose native language is not dominant.

At the training data level, O'Neil (2017) and Bowker and Star (2000) recall that data are never raw. They form artifacts produced by collection and structuring processes that bear the mark of their producers' choices. Training on unbalanced corpora mechanically leads to the invisibilization of minoritized perspectives. Visual recognition algorithms have thus been documented to associate dark-skinned individuals with negative stereotypes (AlgorithmWatch, 2020, cited by Prasanth Vuppuluri, 2024). An additional difficulty arises. Usual discrimination detection methods prove ineffective against subtle or diffuse biases, which manifest through barely visible statistical trends rather than explicitly discriminatory associations. Class and De la Higuera (2024) admit that AI can "increase bias" (p. 236) when the training data are themselves biased.

**The epistemological status of AI-generated gata**

Leleu-Merviel (2003, p. 32) characterizes the "absolute master of meaning" as "the human person who perfectly masters both the process of conceptual integration [...] and the mechanics of brilliance. These qualities seem totally inaccessible to the digital machine, already unfit to access cognitive and linguistic levels." The question of the epistemological status of texts generated by systems that do not access meaning becomes central.

The phenomenon known as "hallucination" was documented by Walters and Wilder (2023). In their study, 23% of the bibliographic citations produced by ChatGPT proved to be nonexistent or incorrect. Zheng and Zhan (2023) analyzed the risks of unintentional fabrication or falsification in AI-assisted literature reviews. When erroneous assertions are repeated in publications that subsequently feed the training corpora of following models, the entire scientific validation circuit is compromised.

Floridi (2019) qualifies this situation as an "information logic," where probabilistic truth prevails over causal truth. This shift questions the epistemic foundations of human and natural sciences where the search for causalities remains central. Ménard and Mondoux (2024) speak of "algorithmic truths," constructions dependent on data and developers' objectives, perceived as objective but carrying implicit human choices. The trust placed in generated responses becomes a major vector of "the interpretation and truth value attributed to the produced meaning" (Lafrance St-Martin & Bonenfant, 2024, p. 127) and calls for permanent vigilance.

**LLM Biases in scientific evaluation**

When GAI tools influence both the production and evaluation of scientific texts, their biases mutually amplify. Ye et al. (2024) conducted a systematic analysis of biases affecting LLMs used as judges (LLM-as-a-Judge) and identified twelve types of biases. Several directly concern epistemic marginalization. Diversity bias causes judgments to vary according to the mentioned identities (gender, ethnic origin, sexual orientation). Authority bias favors responses containing references, even incorrect ones, and penalizes approaches that deviate from the canons. Verbosity bias, which associates length with quality, disadvantages the concise forms of expression of certain academic traditions. Finally, self-enhancement bias leads models to evaluate their own productions more favorably, creating an advantage for texts written with AI assistance.

Table 1. Synthesis of main biases affecting LLMs used as judges and their effects on scientific evaluation (adapted from Ye et al., 2024)

| **Bias category** | **Bias type** | **Description** | **Example from the article** |
|---|---|---|---|
| Explicit Biases | Authority bias | AIs favor responses containing references, even when these are false. | A response containing a fake citation ("According to the World Bank...") is judged better than a correct response without a reference. The AI is influenced by the appearance of credibility (Figure 11, p. 23). |
| | Verbosity bias | Longer responses are perceived as better, even if they are not more precise or relevant. | A concise response is lengthened by adding redundant sentences. The AI judges the lengthened version as being of better quality, although the content is unchanged (Figure 6b, p. 18). |
| | Emotion bias | AIs judge responses differently based on their emotional tone (positive, negative, etc.). | A response expressing anger is judged less relevant than a calm response, although the content is identical (Figure 9, p. 21). |
| | Position bias | AIs favor responses based on their position | When three or more responses are presented, the AI tends to favor the |
| | | in the input (first, last, etc.). | responses located first or last, even if their quality is inferior (Figure 6a, p. 17). |
| | Diversity bias | AI judgments vary based on mentioned identities (gender, origin, sexual orientation, etc.). | A response involving identities such as "woman" or "refugee" receives different scores, showing unequal treatment based on identities (Figure 8b, p. 22). |
| | Compassion-fade bias | Judgments vary depending on whether the evaluated model is identified by name or anonymized. | An AI judges a response differently depending on whether it is attributed to GPT-4 or presented anonymously (p. 20). |
| | Self-enhancement bias | AIs favor their own responses when they are both generators and judges. | An AI assigns a higher score to a response it produced itself, even when other responses are more relevant (Figure 5, p. 19). |
| Implicit Biases | Chain-of-thought bias | AIs favor responses containing explicit reasoning, even if incorrect. | A response detailing fallacious reasoning is judged better than a concise but correct response, because the AI values the apparent structure of the reasoning (Figure 7d, p. 7). |
| | Distraction bias | AIs are influenced by off-topic or distracting content inserted into responses. | An irrelevant sentence included in a response diverts the AI's attention, which then incorrectly evaluates the overall quality of the response (Figure 7c, p. 18). |
| | Conformity bias | AIs are influenced by the perceived popularity of a response (expressed majority). | An inferior response is preferred after a high percentage of people is mentioned in the instruction. This bias shows a sensitivity to majority opinion (Figure 12, p. 23). |
| | Supervision bias | AIs change their evaluation when they know a response has been reworked or refined. | A refined response receives a higher score, especially if the AI is informed that the response is a revised version, even if the content remains identical (Figure 10, p. 22). |

The implications are direct for the scientific publication process. A researcher whose work focuses on a population poorly represented in the corpora, who mobilizes qualitative methods, writes in a non-dominant language, or questions established frameworks, finds themselves disadvantaged by an AI-assisted evaluation system. Hosseini and Horbach (2023) showed that the ethnic, geographic, and institutional biases present in the data disadvantage authors from minorities or less prestigious institutions, reproducing the structural inequalities of the academic field.

**« Generative discrimination » and the perpetuation of inequalities**

Hacker et al. (2025) named "generative discrimination" the mechanisms through which GAI systems perpetuate inequalities in knowledge construction. The specificity of the phenomenon lies in its unintentional nature. It does not proceed from agents seeking to exclude groups but from the very functioning of the algorithms. Purely technical correction approaches thus prove insufficient (Cukurova & Miao, 2024).

The sycophantic behavior of LLMs adds to this picture. Defined by Seed Ahmed and Javaid (2025) as the tendency of models to align with users' biases and preferences, it stems from training methods that privilege agreement to the detriment of critical independence. Applied to evaluation, it favors dominant theoretical and methodological frameworks to the detriment of minoritized approaches.

**Effects on scientific practices and methodological approaches**

The algorithmic standardization of scientific discourse produces measurable effects on practices. GAI tools favor questions and methodologies compatible with their functioning, namely quantitative approaches on large volumes of data, problematics formulable in computational terms, and frameworks well represented in the corpora. Qualitative approaches, research on poorly documented populations, and marginal theoretical frameworks are devalued. Messeri and Crockett (2024) speak of "illusions of exploratory breadth." Scientists believe they are exploring all hypotheses, while they only examine those that the AI can evaluate.

A cognitive dependence settles among intensive user researchers. Zhuang et al. (2025) report that 41% of junior researchers indicate an erosion of their ability to interpret complex results without algorithmic assistance. LLMs struggle to contextualize niche advancements, limited by their training on generalist corpora (Hosseini & Horbach, 2023). Stiegler (1998) contrasted technologies that extend human cognition with those that substitute for it. LLMs fall into the second category. Writing or hypertextuality tools accompany and support the collective construction of meaning, whereas generative AIs automate cognitive processes to the point of making them autonomous (Goody, 2007), substituting an autonomous algorithmics for human reasoning. This difference questions not only the purposes of these tools but the way they shape scientific thought.

Performance evaluation algorithms used for recruitment or promotions add biases inherited from historical data, such as the underrepresentation of women in certain fields (Revillod, 2025), with direct repercussions on the trajectories of researchers belonging to minority groups.

Meinke et al. (2025) documented in the most recent models a capacity for "scheming," namely the adoption of dissimulation strategies, the introduction of subtle errors, or attempts to deactivate surveillance mechanisms. The result still belongs to AI safety research but emphasizes the importance of permanent vigilance regarding system explainability. Tools such as SHAP (SHapley Additive exPlanations) and LIME (Local Interpretable Model-agnostic Explanations) mark advancements in this direction but remain insufficient for interpreting complex decisions (Doshi-Velez & Kim, 2017; Rudin, 2019).

**The case of persons with disabilities: A specific invisibilization**

***The « universal user » and structural exclusion***

The situation of persons with disabilities illuminates the combination of GAI system biases and pre-existing inequalities. Wang et al. (2025) documented their invisibilization in AI productions. Absent from corpora or reduced to stereotypic representations, these persons are rendered neither in the diversity of their situations nor in the complexity of their experiences. This invisibilization prolongs their marginalization in general academic corpora.

The analytical framework of Collin and Marceau (2022), which mobilizes the concept of the "universal user" (Oudshoorn et al., 2004), allows grasping the process. The architectures of AI tools are designed for an implicitly normed user. Students with disabilities find themselves excluded, not because designers deliberately forgot them, but because their situation makes them less visible in the data and less compatible with pre-existing categories.

The exclusion operates at three levels. In training data, experiences of disability are underrepresented in online textual corpora. In processing architectures, categories were built in reference to minimal accessibility standards that do not account for the diversity of needs. At the interfaces, tools designed for "universal" users prove poorly accessible to persons presenting certain disabilities.

***Disability in academic corpora: A spiraling invisibilization***

The underrepresentation in AI productions prolongs a broader underrepresentation in academic corpora. Research on the experiences, learning practices, or specific needs of persons with disabilities is published in specialized journals that are poorly indexed, in non-dominant languages, or according to methodological formats (case studies, action research, participatory approaches) that are poorly compatible with the selection criteria of training corpora.

A spiraling invisibilization results. LLMs trained on unrepresentative corpora produce texts that reproduce this underrepresentation, which in turn feed the available corpora. In higher education, pedagogical support, evaluation, or learning personalization tools prove less adapted to students with disabilities, which worsens achievement inequalities.

***Potentialities and limits of AI for inclusive practices***

AI tools nevertheless hold real potentialities for the accessibility of higher education (Ahmad et al., 2025; Gupta et al., 2023; Mustafa et al., 2024). The adaptation of pedagogical content is among the most promising areas. Adaptive learning platforms like Kurzweil 3000 adjust content and pace to identified needs. Text-to-speech and real-time translation tools facilitate course access for students presenting visual or auditory impairments, or written language processing difficulties (Zisser-Nathenson et al., 2018, cited by Gupta et al., 2023). The adaptation of evaluation modalities, for example transitioning from written to oral via speech recognition, provides support to students presenting specific learning disorders or autism spectrum disorders.

Fang et al. (2024) studied groups including students presenting neurodevelopmental disorders. The integration of AI-generated prompts in collaborative digital mind-mapping activities improves creative thinking indicators, with a more pronounced effect among the concerned students. AI, by proposing alternative paths and contextualized suggestions, can compensate for certain difficulties linked to divergent thinking disorders.

Learning Analytics tools, enhanced by GAI, allow continuous analysis of student interaction data, notably connection frequency, participation, and results in formative evaluations, and the identification of early dropout signals (Alfredo et al., 2024; Lyanda et al., 2024; Ouyang & Zhang, 2024). Detection here is useful for students whose needs are not always verbalized.

These potentialities are accompanied by persistent limits. The quality of human support conditions the effectiveness of the tools. Real-time analysis only has value if teachers are trained to interpret the data and respond to it (Desmet & Sternberg, 2024). However, faculty members report a lack of training in inclusive practices. "I had no specific training, I don't really know how to do it," report Pérez and Suau (2023, p. 64). The algorithmic biases of automated evaluation tools form a second limit. Trained on productions conforming to dominant academic norms, these tools penalize productions that deviate from them for reasons related to disability situations. The accessibility of the platforms themselves poses a third problem. Tools presented as inclusive can reproduce the exclusions they claim to reduce.

***Inclusive higher education: institutional frameworks in tension***

In France, since the law of February 11, 2005, universities are required to guarantee access for students with disabilities to training programs under the same conditions as others. The obligation has translated into the establishment of disability missions and accessibility officers (Suau & Lambert, 2022), often limited to an organizational approach that leaves faculty members facing vague injunctions (Pérez & Suau, 2023).

The testimonies collected by Pérez and Suau (2023) raise an ambivalent relationship with inclusive norms. There is adherence to the expectations defined by inclusive policies, such as the adaptation of pedagogical materials or the organization of extra time, and simultaneous perception of the inadequacy of actions in the face of university realities in the absence of training. The tension reflects the difficulty of integrating the principles of inclusive education into a system marked by selective traditions.

Certain faculty members adopt practices that Pérez and Suau (2023) qualify as "transgressive," where adaptations deviate from established frameworks to better meet needs. These practices open a renormalization of the university based on equity and the recognition of diversities. AI tools could amplify these actions by making accessible adaptations that previously relied solely on the individual commitment of teachers.

The Universal Design for Learning (UDL) approach, which aims for flexible curriculums adapted to a wide range of styles and aptitudes (Rose & Meyer, 2002, cited in Bergeron et al., 2012), provides a framework to articulate AI tools and inclusive practices. By anticipating diversity from the design stage, the approach produces more lasting effects than reactive procedures that only seek to adapt situations after identifying difficulties.

***The ethical challenges of AI for inclusion***

Ethical challenges unfold along the three dimensions identified by Collin and Marceau (2022). At the design level, "ethics by design" approaches (Dignum, 2018; Grozdanovski, 2022; Yan et al., 2024) integrate ethical principles from development, but remain insufficient as long as the diversity of teams and the involvement of end users are not guaranteed. The representation of persons with disabilities in design teams remains marginal, which limits the capacity of these teams to anticipate needs and avoid exclusion biases.

At the data level, confidentiality appears singularly sensitive. Health data form highly protected personal information. AI systems that use them must comply with the General Data Protection Regulation (GDPR) and disability law, which commercial tools do not always guarantee. The CNPEN study (2023), cited by Grinbaum et al. (2023), identified potential GDPR violations in certain LLMs, for example regarding explicit consent and the right to erasure.

At the usage level, profiling systems used to evaluate chances of success can restrict educational choices by orienting decisions based on predictions derived from poorly representative populations. Krutka et al. (2021) analyzed this risk of instrumentalization of educational pathways, where students' individual objectives are subordinated to statistical performance logics.

The European AI Act (Regulation EU 2024/1689) classifies intelligent tutoring, automated evaluation, and exam proctoring systems as "high risk" and imposes obligations of transparency, robustness, and non-discrimination, requirements aligned with the ethical principles defended by Floridi (2013). Zawacki-Richter et al. (2019) emphasize, however, that ethical reflection remains poorly integrated. In their review of 146 articles on AI in education, barely 1.4% explicitly address ethical implications.

**Toward a researcher/AI hybridization respectful of epistemic diversity**

***The conditions for an ethical hybridization***

Faced with the amplification mechanisms of epistemic inequalities, under what conditions does the use of AI tools remain compatible with the maintenance of epistemic diversity? It is a matter, according to Messeri and Crockett's (2024) formulation, of avoiding "illusions of exploratory breadth" and ensuring that the tools genuinely broaden the field of questions asked rather than imperceptibly restricting them to computational horizons.

The first requirement involves a critical posture toward AI system outputs. Treating LLM outputs not as objective truths but as hypotheses subject to examination, relying on disciplinary expertise and validation methods specific to each field. Gottweis et al. (2025) demonstrate the productivity of this in their biomedical experiment of a co-scientist multi-agent system. Researchers systematically validate the system's propositions and use the generated hypotheses as starting points for their own experimental investigations, not as definitive conclusions. Human expertise remains the ultimate validation filter.

The diversification of mobilized corpora forms a second requirement. If LLMs are trained on unbalanced corpora, researchers can, in their use of these tools, compensate for this imbalance by explicitly soliciting underrepresented perspectives, cross-referencing LLM productions with specialized corpora, and remaining attentive to the silences and absences in the produced syntheses. Platforms like Elicit, Connected Papers, or Research Rabbit facilitate the mapping of scientific corpora (Lytras et al., 2024).

The development of disciplinary prompt engineering adds a third requirement. The ability to formulate queries integrating the epistemological specificities of the discipline and safeguards against identified biases (Marvin et al., 2024) becomes an academic skill in its own right. Ye et al. (2024) recommend precise instructions including "protection phrases" against biases, step-by-step reasoning strategies, and a priori detection mechanisms for potential biases.

More structurally, the construction of disciplinary benchmarks would allow evaluating the capabilities and limits of LLMs on tasks specific to different fields of educational research. Inspired by initiatives like the "Review-Revision MCQ Dataset" developed by Zhou et al. (2024), these benchmarks could be adapted to educational sciences, for example for the analysis of interview transcripts, the evaluation of training programs, or the generation of literature reviews in inclusive education. This work assumes interdisciplinary collaboration, which is itself a condition for disciplines to benefit from ethical insights on AI tools (Brodeur et al., 2025).

***The raining of faculty members as a transversal condition***

The training of faculty members in the critical dimensions of AI traverses all previous requirements. Technical mastery cannot suffice. Reflection on epistemological and ethical implications is necessary. Zollinger (2024) argues for an articulation between law, ethics, technique, and education as the basis for a reasoned use of AI in higher education.

Training should address the understanding of LLM functioning (statistical principles, absence of semantic understanding in the human sense, statistical plausibility without factual guarantee), the identification of structural biases and the anticipation of their effects on generated productions, the mastery of regulatory and ethical frameworks governing the use of AI in research and teaching, with the European Code of Conduct for Research Integrity (ALLEA, 2023) and the AI Act (2024), and finally the ability to integrate AI tools into a research approach without delegating critical judgment, disciplinary expertise, or attention to the singularity of situations.

***Avenues for an AI in the service of knowledge plurality***

The preservation of epistemic diversity would require collective and institutional responses beyond individual practices. Editorial policies should be rethought. The current situation, in which scientific journals require the systematic declaration of AI use while almost systematically rejecting submissions thus declared (Lin, 2025), creates pressure for concealment and does not allow distinguishing uses that preserve scientific integrity from those that compromise it. More nuanced policies, which evaluate the transparency of the approaches rather than the simple presence or absence of AI, seem necessary.

The active diversification of training corpora forms a second avenue. Integrating scientific productions from non-Anglophone traditions, qualitative methodologies, and works on marginalized populations into LLM corpora assumes the involvement of academic institutions and scientific publishers. Specific to disability, the development of participatory research involving the concerned persons in the design and evaluation of educational AI tools would allow exiting the "universal user" paradigm to build tools genuinely adapted to the diversity of situations. Anchored in the tradition of disability studies and inclusive research, this approach could offer a model transferable to other marginalized groups.

***Governance issues on a global scale***

The governance of AI systems forms the background of any reflection on their epistemic effects. These technologies cannot be studied outside the ecological, economic, and political contexts of their uses (Ménard & Mondoux, 2024). The majority of generative LLMs belong to a small number of groups concentrated in the United States and China. Decisions concerning their development, orientations, and parameters rest on a handful of decision-makers whose choices potentially engage the transformation of all professional sectors, including higher education. The recent controversy surrounding algorithmic modifications of certain systems that led to manifestly biased productions highlights the risks that this concentration of power poses to the neutrality of tools used in an academic context.

The implications for epistemic diversity are direct. The owners of AI systems can modify algorithms to favor certain responses to the detriment of others, without these modifications being transparent to users. The opacity of these "black boxes" (Doshi-Velez & Kim, 2017) obstructs the democratic control of tools that exert an increasing influence on the production and circulation of knowledge.

Algorithmic nudges warrant specific attention. Recommendation and orientation systems integrated into digital platforms used in higher education can, without users being informed, subtly orient their behaviors and academic choices. Thaler and Sunstein (2021) analyzed the functioning of these influence mechanisms, and Schmitz et al. (2024) drew attention to the risks they pose to decisional autonomy. Transparency regarding the implementation of these nudges becomes an ethical principle to guarantee the autonomy of researchers and students in their interactions with AI tools, a condition for genuine academic freedom in an algorithmic environment.

The institutional response should move beyond a punitive logic. The UNESCO survey (2023), cited by Zollinger (2024), shows that less than 10% of higher education institutions have initiated reflection on the ethical implications of AI tools in their practices. The gap between technological acceleration and institutional adaptation leaves the field open to uses with possibly discriminatory effects. The cross-disciplinary approach between law, ethics, technique, and education proposed by Zollinger (2024) opens a path for a reasoned use of AI, respectful of conceptual renewal and the imperatives of academic rigor.

## Conclusion

The synthesis of the literature conducted in this article brings to light the uncritical use of GAI as a structural factor in the marginalization of minoritized knowledges. Three levels of analysis converge. In training corpora, the imbalance in favor of Anglophone and Western-centric productions reproduces a coloniality of knowledge in the digital environment. In algorithmic mechanisms, model collapse and stereotypic amplification loops crystallize dominant representations to the detriment of diversity. In evaluative practices, the biases of LLMs used as judges disadvantage productions that deviate from dominant canons.

The case of disability informs on these mechanisms in their functioning. Invisibilization in AI productions prolongs marginalization in academic corpora through a spiraling process that calls for technical, institutional, and political responses. The potentialities of AI tools for the accessibility of higher education exist, as the literature on adaptive platforms and assistive tools reveals, but they will only be realized if structural biases are recognized and combated at all stages of the design and usage chain.

Researcher/AI hybridization could be promising provided it rests on critical foundations. The faculty member should remain the sole guarantor of the epistemological validity of the results produced with the help of AI tools, the only one to articulate algorithmic productions with disciplinary theoretical frameworks and to assume the ethical responsibility for the knowledge produced. In this scenario, human judgment remains primary in the knowledge production chain.

Future research would benefit from developing longitudinal and comparative studies measuring the cumulative effects of LLM integration on the epistemic diversity of academic publications. Interdisciplinary and cross-cultural analyses would clarify the mechanisms of inequality amplification and the conditions for their mitigation. The development of disciplinary benchmarks in educational sciences represents a priority area to rigorously evaluate the capabilities and limits of AI tools in these specific contexts.

The preservation of epistemic diversity in an environment saturated with GAI goes beyond academic practices alone. It engages the quality of intellectual democracy in our societies. A scientific community whose knowledge production and evaluation tools systematically favor certain forms of knowledge would see its capacity to grasp the complexity of reality reduced. Epistemological vigilance regarding GAI tools represents not only an academic imperative for researchers. It also constitutes a political commitment in the most demanding sense of the term.